\documentclass{aa}

\usepackage{natbib}
\usepackage{longtable}
\usepackage{lscape}
\usepackage{amsmath}
\usepackage{txfonts}
\usepackage{color}
\usepackage{url}
\usepackage{xspace}

\usepackage{graphicx}
\bibpunct{(}{)}{;}{a}{}{,}

\newcommand{\kev}{keV\xspace}
\newcommand{\ergs}{erg$\,$s$^{-1}$}
\newcommand{\xmm}{\textsl{XMM-Newton}\xspace}
\newcommand{\rosat}{\textsl{ROSAT}\xspace}

\newcommand{\chandra}{\textsl{Chandra}\xspace}

\begin{document}
   \title{An ultraluminous supersoft source with a 4\,hour modulation in NGC~4631}

   \author{S. Carpano
          \inst{1}
          \and
	  A.~M.~T. Pollock
          \inst{1}
          \and
	  A.~R. King
          \inst{2}
          \and
          J. Wilms
          \inst{3}
	  \and
	  M. Ehle
	  \inst{1}}

   \offprints{S. Carpano, e-mail: scarpano@sciops.esa.int}
   \institute{XMM-Newton Science Operations Centre, ESAC, ESA, PO Box 50727, 28080 Madrid, Spain
             \and
	     Department of Physics and Astronomy, University of Leicester, Leicester LE1 7 RH, UK
             \and
	     Dr. Remeis-Observatory, Astronomisches Institut der FAU Erlangen-N\"urnberg, Sternwartstr. 7, 96049 Bamberg, Germany
              }
   
   \date{Submitted: 4 April 2007; Accepted: 6 July 2007}
   
   \abstract{Supersoft X-ray sources (SSSs) are characterised by very low temperatures ($<$100~eV). 
   Classical SSSs have bolometric luminosities in the range of 10$^{36}$--10$^{38}$\,\ergs and are
   modelled with steady nuclear burning of hydrogen on the surfaces of white dwarfs. However, several SSSs have been
   discovered with much higher luminosities. Their nature is still unclear.}{We report the discovery of a
   4\,h modulation for an ultraluminous SSS in the nearby edge-on spiral galaxy NGC~4631, observed with \xmm
   in 2002 June. Temporal and spectral analysis of the source is performed.}{We use a Lomb-Scargle
   periodogram analysis for the period search and evaluate the confidence level using Monte-Carlo
   simulations. We measure the source temperature, flux and luminosity through spectral
   fitting.}{A modulation of $4.2\pm0.4$\,h (3$\sigma$ error) was found for the SSS with a confidence level
   $>$99\%. Besides dips observed in the light curve, the flux  decreased by a factor of 3 within
   $\sim10$\,h. The spectrum can be described with an absorbed blackbody model with $kT\sim67$\,eV. The
   absorbed luminosity in the 0.2--2\,\kev energy band was $2.7\times10^{38}$\,\ergs while the bolometric
   luminosity was a hundred time higher ($3.2\times10^{40}$\,\ergs), making the source one of the most luminous
   of its class, assuming the best fit model is correct.}{This source is another very luminous SSS for 
   which the standard white dwarf interpretation 
   cannot be applied, unless a strong beaming
   factor is considered. A stellar-mass black hole accreting at a super Eddington rate is a more likely interpretation,
   where the excess of accreted
   matter is ejected through a strong optically-thick outflow. 
   The 4\,h modulation could either be an eclipse from the companion star or the consequence of a warped accretion disk.}


\maketitle
%

\section{Introduction}
\label{sec:int}
The X-ray binaries known as supersoft sources (SSSs) are characterised by very soft emission with temperatures $<$100\,eV and bolometric
luminosities exceeding 10$^{36}$\,\ergs. The standard model for such soft and high luminosities was proposed by 
\cite{vandenHeuvel1992} as nuclear burning of hydrogen on the surface of a white dwarf of mass in the range
0.7--1.2\,$M_{\odot}$. On the other hand, if luminosities greatly exceed the Eddington
limit for solar mass compact objects, it is not obvious that this type of model could apply
unless beaming is considered. Several such
very luminous SSSs have been reported in nearby galaxies: two in \object{M~101} \citep{DiStefano2003, Kong2004, Mukai2005}, one in \object{M~51}
\citep{DiStefano2003}, one in \object{M~81} \citep{Swartz2002}, one in \object{the Antennae} \citep{Fabbiano2003} and two in 
\object{NGC~300} \citep{Kong2003,Carpano2006}.

In one of these SSSs, a short period of 5.4\,h has been reported: the source located in the
face-on spiral galaxy NGC~300, which was discussed by \cite{Kong2003} and \cite{Carpano2006}. The modulation
was only present during a single \xmm observation but was not visible 6 days earlier, when the source was a
few times brighter. In this Letter, we report the discovery of an even  more luminous SSS with a slightly shorter modulation, 
located in the nearby edge-on galaxy \object{NGC~4631}.

NGC~4631 is a SBc/d type galaxy located at an assumed distance of 7.5\,Mpc \citep{Golla1994}, with  low galactic foreground absorption 
\citep[$N_\text{H}=1.2\times10^{20}\,\text{cm}^{-2}$;][]{Dickey1990}. A study of the giant diffuse X-ray emitting
corona around the galaxy was performed by \cite{Wang2001} with \chandra and by \cite{Tullmann2006}
with \xmm. The point-source population
viewed by \rosat was studied by \cite{Vogler1996} and \cite{Read1997}. They reported the detection of 7 sources, 
including the supersoft source described in this Letter. Spectral fitting of this SSS using a bremsstrahlung model 
\citep{Read1997} gave a temperature of 0.1\,\kev with high 
($8\times10^{21}\,\text{cm}^{-2}$) intrinsic absorption; the corresponding absorbed luminosity in the 0.1--2.0\,\kev energy
band was of $1.9\times10^{38}$\,\ergs. In the long-term light curve \citep{Vogler1996} the
source was not visible in two of six data sets although, because the source flux was close to the detection limit 
in the other four, time variability could not be established.

In this Letter, we report the discovery of a 4\,h modulation for the ultraluminous SSS in NGC~4631. The 
remainder of the Letter is organised as follows. Section~\ref{sec:obs} describes the \xmm
observation and data reduction. In Sect.~\ref{sec:time}, we present
timing and spectral analysis of the SSS, reserving discussion for
Sect.~\ref{sec:conc}.  

\section{Observation and data reduction}
\label{sec:obs}
NGC~4631 was observed by \xmm on 2002 June 8 for 55\,ksec. The  EPIC-MOS
\citep{Turner} and EPIC-pn \citep{Strueder} cameras were operated in
full frame mode with the medium MOS filter and the thin pn filter. 
After screening the MOS data for proton flares using standard 
procedures\footnote{\url{http://xmm.esac.esa.int/external/xmm_user_support/documentation/sas_usg/USG/}},
a total of 45 and 37\,ksec of low-background emission remained for the MOS and pn, respectively. 

Using the \xmm Software Analysis System (SAS) \texttt{edetect\_chain} task, which performs maximum-likelihood source detection, the SSS
was detected with a maximum likelihood of 1919. The best-fit coordinates were:
$\alpha_\text{J2000}=12^\text{h}42^\text{m} 16\fs{}1$ and
$\delta_\text{J2000}=+32^\circ 32' 49\farcs 5$ with a statistical error of $0\farcs 3$. 

\section{Timing and spectral analysis of the SSS}
\label{sec:time}
\begin{figure}
  \resizebox{\hsize}{!}{\includegraphics{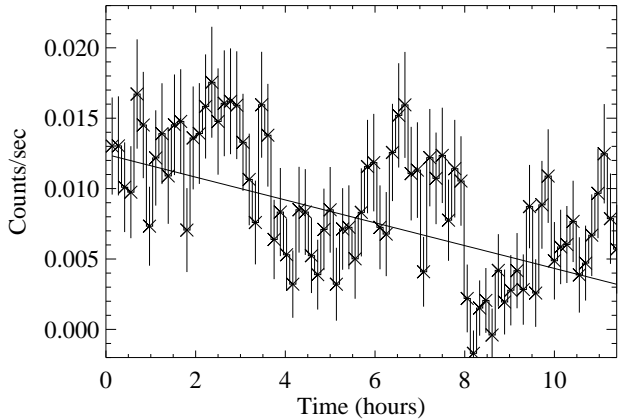}}
 \caption{Mean \xmm EPIC-MOS and pn background-subtracted light curve of the SSS in NGC~4631.  
 Error bars are at 1 $\sigma$.
 The straight line shows the best linear fit to the lightcurve.}
 \label{fig:light}
\end{figure}

The combined EPIC-MOS and pn background-subtracted lightcurve is shown in Fig.~\ref{fig:light}, with a time bin
size of 500\,sec. Times are given in hours from the start of the observation.
Periods of high background at the end of the observation have been excluded from the data.
Two dips are apparent in the lightcurve in addition to a long-term decrease by a factor of
3 (Fig.~\ref{fig:light}, straight line).

\begin{figure}
  \resizebox{\hsize}{!}{\includegraphics{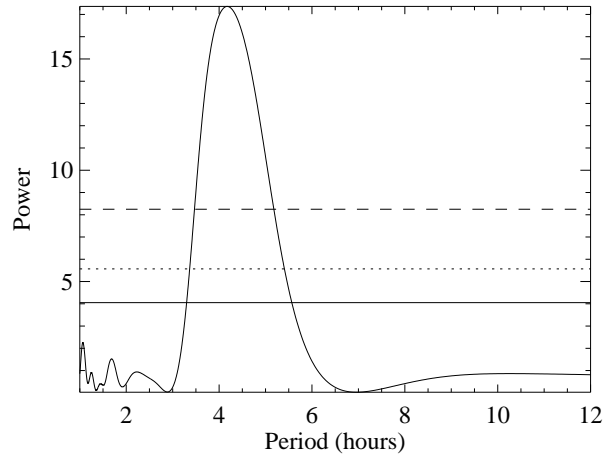}}
 \caption{Search for periodicities in the \xmm light curve of the NGC~4631 SSS using
 Lomb-Scargle periodogram analysis, after removing the decreasing flux trend. The full, dotted and dashed lines represent the 68\%, 90\% and
 99\% confidence levels, respectively, as determined from Monte Carlo simulations.}
 \label{fig:period}
\end{figure}

We performed a period search with Lomb-Scargle periodogram analysis
\citep{Lomb1976,Scargle1982} after having removed the
general trend of the decreasing flux.
Confidence levels were measured by means of Monte Carlo simulations
assuming a null hypothesis of white noise.
Results  are shown in Fig.~\ref{fig:period}. We found that the
4.17\,h period is significant at $>99\%$ confidence. 
This modulation, although less significant, is also present in the MOS data alone.
We estimated the uncertainty of the period by fitting a 
sine function using the IDL\footnote{\url{http://www.ittvis.com/idl}} task \texttt{curvefit} to the
detrended light curve.   The best-fit period by this method was
$4.11^{+0.40}_{-0.34}$\,h, close to the value from the
Lomb-Scargle analysis. The errors quoted are at the $3\sigma$
level. 

\begin{figure}
  \resizebox{\hsize}{!}{\includegraphics{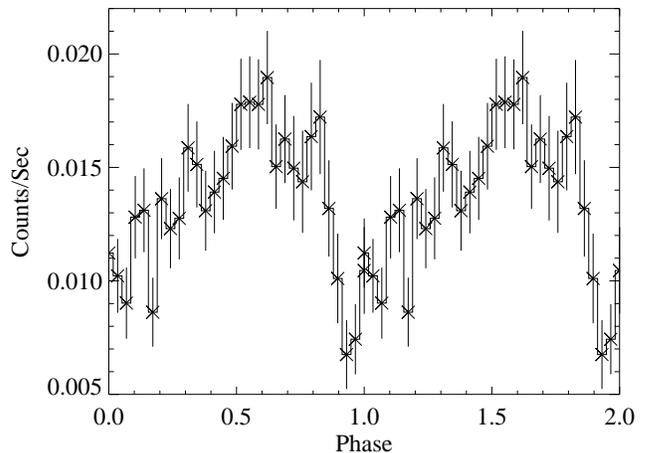}}
 \caption{\xmm light curve of the NGC~4631 SSS folded at 4.17\,h using 30 bins, after removing the decreasing flux trend.}
 \label{fig:fold}
\end{figure}

The corresponding light curve folded at 4.17\,h is shown in
Fig.~\ref{fig:fold}. The shape is slightly asymmetric, the flux change before the dip
sharper than afterwards. The dip is also relatively narrow.

\begin{figure}
  \resizebox{\hsize}{!}{\includegraphics[bb=113 44 563 708,clip=true,angle=-90]{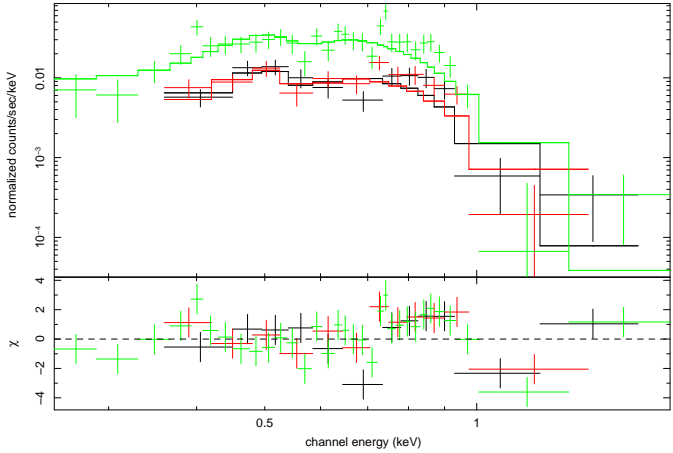}}
 \caption{\xmm EPIC-pn and MOS spectra of the SSS in NGC~4361 and the best-fit spectral
   model with the apparent line emission. Bottom: residuals expressed in $\sigma$.}
 \label{fig:spec}
\end{figure}

Fig.~\ref{fig:spec} shows the pn and MOS spectra of the SSS. The data are
binned to have at least 25 counts in each energy bin. Almost no flux is visible above
2\,\kev. We first tried to model the data with an absorbed single blackbody, yielding 
$\chi^2_{\nu}=1.95$.  From the residuals shown in
Fig.~\ref{fig:spec}, some excess around 0.85\,\kev is observed. Adding
a Gaussian line significantly improved the fit, resulting in a reduced
$\chi^2_{\nu}=1.22$ for $\nu=53$ degrees of freedom. 
The absorbed bremsstrahlung model used by \cite{Read1997}, with or without a Gaussian line, does not describe the data
well ($\chi^2_{\nu}=3$).
We also tried to fit a black body with a power-law to describe the continuum but the normalisation
constant of this last component is very small and spectral parameters 
as well as luminosities do not change significantly (less than a factor of 2). Alternatively to the gaussian 
emission line at 0.8\,\kev, an absorption edge could also
describe the spectrum ($\chi^2_{\nu}=1.35$). Unfortunately due to the low statistic of the data above 1\,\kev,
the absoprtion depth parameter cannot be constrained.
Absorption edges have been reported by \cite{Swartz2002} and \cite{Kong2004} for the luminous SSSs in M~81
and M~101, respectively, indicating the presence of a warm absorber in the vicinity of the X-ray source.
The best-fit parameters of the absorbed combined blackbody and single
Gaussian-line model are shown in Table~\ref{tab:spec_fit}. $N_\text{H}$ is the
equivalent column density of neutral hydrogen, $kT$ the temperature, $E_{\text{L}}$ the energy of the line and
$\sigma_{\text{L}}$ its width. The
corresponding 0.2--2 \,\kev flux, absorbed, 
unabsorbed and bolometric luminosities are shown in the last four
rows. Uncertainties are given at the 90\% confidence level, except
for the bolometric luminosity at 99\%. As the source was close to a pn CCD gap, the
flux and luminosities were calculated from the MOS data alone.

.

Our data modelling shows the source to have a very high intrinsic
absorption column giving a bolometric
luminosity that makes the SSS one of the most luminous of its class.
Such high absorption could be due to a strong intrinsic absorption in the vicinity of the emitting
source or due to gas of the hosting galaxy since it is edge-on oriented.
The source is also highly variable.
The object has also been observed once with \chandra, about two years before the \xmm observation,
on 2000 April 16 for 59\,ksec when 31 net
counts were detected. Converting this rate to flux using 
\texttt{WebPIMMS}\footnote{\url{http://heasarc.gsfc.nasa.gov/Tools/w3pimms.html}}
with the spectral parameters of
Table~\ref{tab:spec_fit},
we estimate that the mean observed luminosity at that time
in the 0.2--2\,\kev energy band was $1.7\times10^{37}$\,\ergs.
This is about a factor of 10 lower than the \xmm measurement 
two years later and the earlier
\rosat value \citep{Read1997}.

\begin{table}
 \caption{Results of the spectral fits for the SSS in NGC~4361, using an
   absorbed blackbody model and a Gaussian line
   (\texttt{phabs*(bbody+gauss)}, in XSPEC). } 
  \label{tab:spec_fit}
 \begin{tabular}{ll}
 \hline
 Parameter & Value (units)\\[3pt]
 \hline
 $N_\text{H}$& 4.94$^{+1.61}_{-1.64}\times 10^{21} (\text{cm}^{-2})$\\[3pt]
 $kT$ & 67$^{+5}_{-15}$ (eV)\\[3pt]
 E$_{\text{\,L}}$& 0.80$^{+0.03}_{-0.05}$ (\kev)\\[3pt]
 $\sigma_{\text{\,L}}$ &  0.07$^{+0.04}_{-0.02}$ (\kev)\\[3pt]
 $F_{0.2-2\,\text{\kev}}$& 3.97$^{+4.76}_{-1.57}\times 10^{-14}$ (\,erg$\,$s$^{-1}$\,cm$^{-2}$) \\[3pt]
 $L_{0.2-2\,\text{\kev}}^{\text{obs}}$& 2.67$^{+3.20}_{-1.06}\times 10^{38}$ (\,\ergs) \\[3pt]
 $L_{0.2-2\,\text{\kev}}^{\text{unabs}}$& 1.99$^{+0.59}_{-1.27}\times 10^{40}$ (\,\ergs) \\[3pt]
 $L_{\text{bol}}$ & 3.22$^{+1.35}_{-2.88}\times 10^{40}$ (\,\ergs) \\[3pt]
 \hline
   \end{tabular}
\end{table}


\section{Discussion}
\label{sec:conc}
The SSS reported in this Letter is one of the most luminous  of its class. Besides the bright and soft X-ray
emission, the source is particular for its 4\,h modulation superimposed on a decreasing flux trend during 
the \xmm observation. On
only one other occasion, for a SSS in NGC~300 \citep{Kong2003}, has a short modulation been reported
for such a bright SSS, of 5.4\,h in that case,
in the one of the two \xmm observations in which the source was fainter.
A discussion of the source has also been reported by \cite{Carpano2006}.

The high luminosity ($>10^{40}$\,\ergs) of the NGC~4361 SSS, which exceeds the Eddington luminosity for a white dwarf or a neutron star
by a factor of 100, might be due to a strong wind or outflow of a stellar-mass black hole accreting at a
super Eddington rate. This idea has also been suggested by \cite{Fabbiano2003} and \cite{Mukai2005} for the
ultraluminous SSSs in the Antennae and M~101, respectively. The model has been reconsidered in more detail
by \cite{King2003}, as summarised below. Other less convincing models in the literature,
as summarised by \cite{Carpano2006}, have variously involved strong beaming from steady nuclear burning of
hydrogen at the surface of a white dwarf; accretion onto an intermediate-mass black hole; or soft emission of an accretion disk
observed at high inclination.

\cite{King2003} suggested that a black hole accreting at super-Eddington rate, ejects the excess
matter through a strong wind or outflow that is likely to be optically thick to electron scattering
for $\dot{M}_\text{out}\sim\dot{M}_\text{Edd}$ with a photospheric radius of the order of few tens of the
Schwarzschild radius of the black hole. The outflow is radial and occupies a double cone of solid angle 
$4 \pi b$, where the beaming factor $b$ can be either $\sim1$ or $\ll1$. 
The emission line present in the spectrum of the SSS in NGC~4631, which is 
likely to be composed of several unresolved lines, or alternatively the absorption edge,
might come from optically thin parts of the matter outflow. These could 
either be near the edges of the conical outflow, or the leading edge of 
the outflow if this is time-dependent.
The temperature resulting from the black
body emission is \citep{King2003}:
\begin{equation}
T_\text{eff}=1\times10^5 g^{-1/4} \dot{M}_1^{-1} M_8^{3/4}\,\text{K}
\label{equ:teff}
\end{equation}
where $g(b)=1/b$ or $1/(2b^{1/2})$ (for $b\sim1$ or $b\ll1$), 
$\dot{M}_1=\dot{M}_{\text{out}}/(1\,\text{M}_{\odot}\text{yr}^{-1})$,
and $M_8=M/10^8\,\text{M}_\odot$, and $M$ is the mass of the accretor.  

Assuming the source luminosity is close to the Eddington limit, $\dot{M}_\text{out}\sim\dot{M}=L/(\eta
\text{c}^2)$, where $L$ is the luminosity and $\eta$ the radiative efficiency, we are able to estimate the mass of
the black hole
\begin{equation}
  M=0.27 \left[\frac{g^{1/4}}{\eta}\right]^{4/3} \text{M}_\odot
\end{equation}
As $g^{1/4}$ is of order unity unless $b\lesssim0.1$, for a typical value of $\eta\sim0.1$, the
bolometric luminosity of $3.2\times10^{40}$\,\ergs gives a
corresponding black-hole mass of 6\,$\text{M}_\odot$ and an accretion rate of
$5.6\times\,10^{-7}\,\text{M}_\odot\text{yr}^{-1}$. Such an accretion rate can easily be achieved 
either by thermal-time-scale mass transfer after a high-mass companion has filled its Roche Lobe or during
outbursts of soft X-ray transients \citep{King2002}.

The 4\,h modulation could be caused either by an eclipse from the companion star or by the presence of a warped disk. This latter hypothesis is
driven by the fact that the 5.4\,h modulation in the SSS of NGC~300 was observed in only one of two \xmm observations. 
\cite{Barnard2006} have recently reported that the X-ray binary \object{Bo~158} in \object{M~31}, shows regular dips at 2.78\,h but not in all of their
observations. These authors suggested this dipping to be caused by an elongated precessing disk. 

To summarise, the ultraluminous SSS in NGC~4631 reported in this Letter may be an X-ray binary system with a
short ($\lesssim$10\,h) orbital period, where the high mass transfer rate is either due to a massive companion filling its Roche Lobe 
or because the system is in outburst. When matter is
accreted above the Eddington rate, the excess matter is ejected via a strong, optically-thick outflow
in which case
a mass of 6\,$\text{M}_\odot$ can be estimated for the black-hole companion.

\begin{acknowledgements}
  This paper is based on observations obtained with \textsl{XMM-Newton}, an ESA
  science mission with instruments and contributions directly funded
  by ESA Member States and NASA.
\end{acknowledgements}


\begin{thebibliography}{30}
\expandafter\ifx\csname natexlab\endcsname\relax\def\natexlab#1{#1}\fi

\bibitem[{Barnard}  {et~al.}(2006)]{Barnard2006}
 {Barnard}, R., {Foulkes}, S.~B.,  {Haswell}, C.~A., {Kolb}, U., {Osborne}, J.~P. 
 \& {Murray}, J.~R. 2006, \mnras, 366, 287

\bibitem[{{Carpano} {et~al.}(2006){Carpano}, {Wilms}, {Schirmer}, \&
  {Kendziorra}}]{Carpano2006}
{Carpano}, S., {Wilms}, J., {Schirmer}, M., \& {Kendziorra}, E. 2006, \aap,
  458, 747

\bibitem[{{Dickey} \& {Lockman}(1990)}]{Dickey1990}
{Dickey}, J.~M. \& {Lockman}, F.~J. 1990, \araa, 28, 215

\bibitem[{Di Stefano} \& {Kong} (2003)]{DiStefano2003}
{Di Stefano}, R. \& {Kong}, A.~K.~H. 2003, \apj, 592, 884
	
\bibitem[{Fabbiano}  {et~al.}(2003)]{Fabbiano2003}
{Fabbiano}, G., {King}, A.~R., {Zezas}, A., {Ponman}, T.~J., {Rots}, A. \&
{Schweizer}, F. 2003, \apj, 591, 843

\bibitem[{Golla} \& {Hummel}(1994)]{Golla1994}
{Golla}, G. \& {Hummel}, E. 1994, \aap, 284, 777

\bibitem[{{King}(2002)}]{King2002}
{King}, A.~R. 2002 \mnras, 355, L13

\bibitem[{{King} \& {Pounds}(2003)}]{King2003}
{King}, A.~R. \& {Pounds}, K.~A. 2003, \mnras, 345, 657
	
\bibitem[{{Kong} \& {Di Stefano}(2003)}]{Kong2003}
{Kong}, A.~K.~H. \& {Di Stefano}, R. 2003, \apjl, 590, L13

\bibitem[{{Kong} {et~al.}(2004)}]{Kong2004}
{Kong}, A.~K.~H., {Di Stefano}, R. \& {Yuan}, F. 2004, \apjl, 617, L49

\bibitem[{Lomb}(1976)]{Lomb1976}
{Lomb}, N.~R, 1976, \apss, 39, 447

\bibitem[{{Mukai} {et~al.}(2005)}]{Mukai2005}
{Mukai}, K., {Still}, M., {Corbet}, R.~H.~D., {Kuntz}, K.~D. \& {Barnard}, R.
2005, \apj, 634, 1085

\bibitem[{{Read} {et~al.}(1997) {Read},  {Ponman}, \& {Strickland}}]{Read1997}
{Read}, A.~M., {Ponman}, T.~J. \& {Strickland}, D.~K. 1997, \mnras, 286, 626

\bibitem[{Scargle}(1982)]{Scargle1982}
{Scargle}, J.~D., 1982, 263, 835


\bibitem[{{Str{\" u}der} {et~al.}(2001){Str{\" u}der}, {Briel}, {Dennerl},
  {Hartmann}, {Kendziorra}, {Meidinger}, {Pfeffermann}, {Reppin}, {Aschenbach},
  {Bornemann}, {Br{\" a}uninger}, {Burkert}, {Elender}, {Freyberg}, {Haberl},
  {Hartner}, {Heuschmann}, {Hippmann}, {Kastelic}, {Kemmer}, {Kettenring},
  {Kink}, {Krause}, {M{\" u}ller}, {Oppitz}, {Pietsch}, {Popp}, {Predehl},
  {Read}, {Stephan}, {St{\" o}tter}, {Tr{\" u}mper}, {Holl}, {Kemmer},
  {Soltau}, {St{\" o}tter}, {Weber}, {Weichert}, {von Zanthier},
  {Carathanassis}, {Lutz}, {Richter}, {Solc}, {B{\" o}ttcher}, {Kuster},
  {Staubert}, {Abbey}, {Holland}, {Turner}, {Balasini}, {Bignami}, {La
  Palombara}, {Villa}, {Buttler}, {Gianini}, {Lain{\' e}}, {Lumb}, \&
  {Dhez}}]{Strueder}
{Str{\" u}der}, L., {Briel}, U., {Dennerl}, K., {et~al.} 2001, \aap, 365, L18

\bibitem[{Swartz} {et~al.}(2002)]{Swartz2002}
{Swartz}, D.~A., {Ghosh}, K.~K., {Suleimanov}, V., {Tennant}, A.~F. \& {Wu}, K.
2002, \apj, 574, 382

\bibitem[{T{\"u}llmann} {et~al.}(2006)]{Tullmann2006}
{T{\"u}llmann}, R., {Breitschwerdt}, D., {Rossa}, J., {Pietsch}, W. \& {Dettmar}, R.-J. 2006, \aap, 457, 779
 

\bibitem[{{Turner} {et~al.}(2001){Turner}, {Abbey}, {Arnaud}, {Balasini},
  {Barbera}, {Belsole}, {Bennie}, {Bernard}, {Bignami}, {Boer}, {Briel},
  {Butler}, {Cara}, {Chabaud}, {Cole}, {Collura}, {Conte}, {Cros}, {Denby},
  {Dhez}, {Di Coco}, {Dowson}, {Ferrando}, {Ghizzardi}, {Gianotti}, {Goodall},
  {Gretton}, {Griffiths}, {Hainaut}, {Hochedez}, {Holland}, {Jourdain},
  {Kendziorra}, {Lagostina}, {Laine}, {La Palombara}, {Lortholary}, {Lumb},
  {Marty}, {Molendi}, {Pigot}, {Poindron}, {Pounds}, {Reeves}, {Reppin},
  {Rothenflug}, {Salvetat}, {Sauvageot}, {Schmitt}, {Sembay}, {Short},
  {Spragg}, {Stephen}, {Str{\" u}der}, {Tiengo}, {Trifoglio}, {Tr{\" u}mper},
  {Vercellone}, {Vigroux}, {Villa}, {Ward}, {Whitehead}, \& {Zonca}}]{Turner}
{Turner}, M.~J.~L., {Abbey}, A., {Arnaud}, M., {et~al.} 2001, \aap, 365, L27


\bibitem[{{van den Heuvel} {et~al.}(1992) {van den Heuvel}, 
{Bhattacharya}, {Nomoto},\& {Rappaport}}]{vandenHeuvel1992}
{van den Heuvel}, E.~P.~J., {Bhattacharya}, D., {Nomoto}, K. \& {Rappaport}, S.~A.
1992, \aap, 262, 97

\bibitem[{Vogler} \& {Pietsch}(1996)]{Vogler1996}
{Vogler}, A. \& {Pietsch}, W. 1996, \aap, 311, 35

\bibitem[{Wang} {et~al.}(2001)]{Wang2001}
{Wang}, Q.~D., {Immler}, S., {Walterbos}, R., {Lauroesch}, J.~T. \& {Breitschwerdt}, D. 2001, \apjl, 555, L99

\end{thebibliography}
\end{document}